\documentclass[aps,prd,nofootinbib,showpacs,preprintnumbers,amssymb, twocolumn,11pt]{revtex4-1} 
\usepackage{graphicx}
\usepackage{epsfig}
\usepackage{dcolumn}
\usepackage{bm}
\usepackage{amsmath}
\graphicspath{{./Figures/}}

\newcommand{\brf}{\it B}  \newcommand{\timesdot}{\cdot}

\begin{document}

\preprint{FERMILAB-PUB-13-185-T}

\title{Coupling--mass mapping of di-jet peak searches}

\author{Bogdan A. Dobrescu}
\email{bdob@fnal.gov}
\affiliation{Theoretical Physics Department, Fermilab,
Batavia, IL 60510, USA}
\author{Felix Yu}
\email{felixyu@fnal.gov}
\affiliation{Theoretical Physics Department, Fermilab, 
Batavia, IL 60510, USA}

\date{June 11, 2013; Revised September 10, 2014}

\begin{abstract}
We study hypothetical gauge bosons that may produce dijet resonances
at the LHC.  Simple renormalizable models include leptophobic
$Z^\prime$ bosons or colorons that have flavor-independent couplings
and decay into a color-singlet or -octet quark-antiquark pair,
respectively.  We present the experimental results on dijet resonances
at hadron colliders as limits in the coupling-versus-mass plane of a
gauge boson associated with baryon number.  This theoretical framework
facilitates a direct comparison of dijet resonance searches performed
at different center-of-mass energies or at different colliders.
\end{abstract}


\maketitle

\section{Introduction}
\label{sec:Introduction}

If a new particle is produced in the $s$-channel at hadron colliders,
then it can decay into a pair of hadronic jets (``dijet"). The
invariant mass distribution of the dijet exhibits a peak at (or
slightly below) the mass of the new particle \cite{Han:2010rf,
  Harris:2011bh}.  In many models, particles produced in the
$s$-channel can decay into leptons or other final states with low
backgrounds.  If the branching fractions of those final states are
small enough, however, the dijet resonance searches may provide the
simplest way of discovering the new particles.

Searches for narrow dijet resonances at hadron colliders have been
performed over the last three decades by the UA2 \cite{Alitti:1990kw,
  Alitti:1993pn} and UA1 \cite{Albajar:1988rs} experiments at the SPS
collider, the CDF \cite{Abe:1989gz, Abe:1993it, Abe:1995jz,
  Abe:1997hm, Aaltonen:2008dn} and D0 \cite{Abazov:2003tj} experiments
at the Tevatron, and the ATLAS \cite{Aad:2010ae, Aad:2011aj,
  ATLAS-CONF-2011-081, ATLAS-CONF-2011-095, Aad:2011fq, ATLAS:2012pu,
  ATLAS-CONF-2012-088, ATLAS-CONF-2012-148} and CMS
\cite{Khachatryan:2010jd, Chatrchyan:2011ns, CMS:2012yf,
  CMS-PAS-EXO-11-094, Chatrchyan:2013qha, CMS-PAS-EXO-12-059}
experiments at the LHC.  The results are traditionally presented as
limits on an effective rate (defined as cross section times branching
fraction times acceptance) to produce a resonance as a function of its
mass.  While this procedure has the advantage of being rather
model-independent, it complicates the comparison of experimental results
with theoretical models.  

The acceptance, in particular, requires the computation of the
probability for the two jets to be observed in a certain geometric
region of the detector. This can be done analytically given the
differential cross section of the signal and the kinematic cuts, as
long as effects arising from showering, from an assumed Gaussian
signal (in the case of ATLAS~\cite{Aad:2011fq}), or from a mismatch
between partons and analysis-level wide jets are
negligible. Otherwise, it is necessary to perform a simulation based
on the jet selection criteria used by the experimental analyses.

The effective rate procedure also precludes a comparison of the limits
set in $p\bar{p}$ collisions (at the SPS and the Tevatron) with those
from $pp$ collisions (at the LHC).  Even for a particular collider, it
is hard to compare the limits set during runs of different energies,
because the cross section grows with the center-of-mass energy
($\sqrt{s}\, $) for a fixed resonance mass.  A naive hope is that
limits set at the larger $\sqrt{s}$ and with larger integrated
luminosity would supersede previous limits.  The situation is not so
straightforward because the backgrounds also increase so that the
trigger thresholds for a jet-only final state need to be increased. As
a result, the sensitivity to lighter resonances can be better in the
runs using lower luminosity or lower energy. For example, the ATLAS
dijet limits from $\sqrt{s} = 7$ TeV start at a resonance mass that
has increased with luminosity from 0.3 TeV~\cite{Aad:2010ae} to 0.6
TeV~\cite{Aad:2011aj} to 0.9 TeV~\cite{Aad:2011fq} to 1
TeV~\cite{ATLAS:2012pu}, and those from $\sqrt{s} = 8$ TeV start at a
mass 1.5 TeV~\cite{ATLAS-CONF-2012-088, ATLAS-CONF-2012-148}.

In this article we explore a unified presentation of the dijet limits
in a coupling-versus-mass plot.  The mass and coupling refer to a
certain hypothetical particle, of a given spin and $SU(3)_c\times
SU(2)_W \times U(1)_Y$ charges, which is produced in the $s$-channel
at hadron colliders and decays into a pair of jets.  This is by no
means a substitute for the effective rate plots, as it is more
model-dependent.  The coupling-versus-mass plot, however, has the
advantage of allowing simple comparisons of searches performed at
different luminosities, experiments, $\sqrt{s}$ or colliders.
Furthermore, it provides a measure of how stringent the limits are
given some natural ranges for the physical parameters.

Specifically, we consider an electrically-neutral spin-1 particle that
couples in a flavor-universal way to the SM quark-antiquark pairs and
is leptophobic, {\it i.e.}, its tree-level couplings to SM leptons
vanish.  This is well motivated by the following arguments.  In many
theories beyond the SM, there are particles that can be produced from
a quark-antiquark initial state and lead to a dijet resonance with
large rates. By contrast, both the gluon-gluon (as in the case of the
Higgs boson) and quark-gluon initial states require a loop to produce
an $s$-channel resonance, so that the signal is typically too small
(at least in perturbative theories) to compete with the dijet
background.  The quark-quark initial state could lead to an
$s$-channel resonance if there is a di-quark scalar, but in that case
flavor-changing processes typically impose strong constraints on its
mass and couplings (these are relaxed in the case of the color-sextet,
hypercharge-4/3 di-quark \cite{Bauer:2009cc}).

Electroweak symmetry suppresses the coupling of spin-0 particles to
first generation $q\bar{q}$ pairs (an exception is the color-octet
weak-doublet scalar \cite{Manohar:2006ga}, but in that case there are
strong flavor constraints). Leptophobic spin-2 particles, although
possible, require much more complicated UV completions.

Including a spin-1 particle coupled to first generation quarks is more
straightforward.  Large flavor effects are avoided if its quark
couplings are generation-independent. Moreover, the spin-1 particle
should be associated with a spontaneously-broken gauge symmetry
(unless the particle is a bound state whose compositeness scale is
near its mass), and the cancellation of various gauge anomalies is
more easily achieved for equal couplings to up- and down-type quarks.
Although some of the above arguments can be evaded (for example, with
a more complicated fermion sector that is anomaly-free), a
flavor-universal gauge boson appears to be the simplest origin of a
dijet peak.  In order to couple to the SM quarks, the heavy gauge
boson must be either a singlet or an octet under the $SU(3)_c$ color
group.

In the case of the color singlet (a $Z^\prime$ boson), the dijet
channel can be the discovery mode only if the $Z^\prime$ is nearly
leptophobic (for an early model, see \cite{Babu:1996vt}) and its
decays into Higgs states~\cite{Georgi:1996ei} or vectorlike
leptons~\cite{Rosner:1996eb} are suppressed.  We consider $Z^\prime$
bosons whose tree-level leptonic and Higgs couplings vanish, implying
that the gauge charges are proportional to the baryon number. The
corresponding $U(1)_B$ symmetry is anomalous in the SM, but we will
show that it is anomaly-free in the presence of a few vector-like
quarks (the simplest charge assignment has been discussed in
\cite{FileviezPerez:2011pt}).

A color-octet gauge boson, referred to as the
coloron~\cite{Hill:1991at}, is associated with a $SU(3)_1\times
SU(3)_2$ extension of QCD, and is automatically leptophobic.  The
coloron, in the case of flavor-universal
couplings~\cite{Chivukula:1996yr}, can arise from a simple
renormalizable extension of the SM~\cite{Bai:2010dj}. Although its
low-energy effects are usually negligible (in contrast to the case of
flavor-dependent couplings \cite{Chivukula:2013kw}), the coloron can
modify Higgs production via gluon-fusion \cite{Kumar:2012ww}.

In Section~\ref{sec:Models} we present some simple renormalizable
models that include a $Z^\prime$ boson coupled to baryon number
($Z^\prime_B$) or a coloron ($G^\prime$).  In
Section~\ref{sec:Searches} we use the existing experimental limits on
the effective rate to derive the limits in the coupling--mass
plane for $Z^\prime_B$, and also for $G^\prime$.  Section IV includes
our conclusions.

\section{Models of dijet resonances}
\label{sec:Models}

In this section we present some renormalizable models of spin-1
particles that are either color singlets ($Z^\prime$) or octets
(coloron) and couple to quark-antiquark pairs.

\subsection{$Z^\prime$ coupled to baryon number}
\label{subsec:Models}

Each coupling of a $Z^\prime$ boson to a quark or a lepton is in
principle a free parameter.  In practice, though, there are various
theoretical and phenomenological constraints on these couplings.
Massive spin-1 particles, such as $Z^\prime$ bosons, must be either
bound states or else be associated with a spontaneously-broken gauge
symmetry, the simplest case being a new $U(1)$ group.  At hadron
colliders, in order to discover a dijet resonance before a dilepton
resonance, some of the $Z^\prime$ couplings to quarks must be more
than an order of magnitude larger than all $Z^\prime$ couplings to
leptons.

In the limit where the tree-level couplings to leptons vanish, the
case of a leptophobic $Z^\prime$, there are severe constraints from
the requirement of anomaly cancellations. These form a set of linear,
quadratic and cubic equations in the $U(1)$ charges, which must have
as a solution a set of commensurate numbers ({\it i.e.}, a set of
integers upon a rescaling of the gauge coupling).  Despite the
intricacies of cubic equations for integers, it can be proven that
there is always a solution in the presence of a certain set of
fermions which are vectorlike with respect to the SM gauge group and
chiral under $U(1)$ \cite{Batra:2005rh}.  To have a viable model,
however, the number of new fermions cannot be too large, and their
properties must avoid various phenomenological constraints.

Let us construct some viable models where the $U(1)_B$ symmetry
associated with baryon number is gauged, {\it i.e.}, all SM quarks
have $U(1)_B$ charge 1/3, while all SM leptons and bosons have charge
0. This choice is convenient because the SM mechanism for generating
quark masses is not affected by the additional gauge symmetry, and
furthermore the $Z^\prime$ couplings to quarks are flavor blind.

We construct a class of explicit models of this type that include $n$
sets of vectorlike quarks (color-triplets) transforming under
$SU(2)_W$ as doublets, $Q^k$, or singlets, $U^k, D^k$; here $k = 1,
\ldots , n$ labels their flavor.  Although these new fermions do not
introduce anomalies involving only SM gauge groups, the $U(1)_B$
charges of the vectorlike quarks are restricted by anomaly
cancellation.  The $[SU(2)_W]^2 U(1)_B$, $[SU(3)_c]^2 U(1)_B$ and
$[U(1)_Y]^2U(1)_B$ anomalies cancel only if
\begin{eqnarray}  \hspace*{-6mm}
z_{U_L} \! - z_{U_R}  =   z_{D_L} \! - z_{D_R}  = - z_{Q_L} \! + z_{Q_R}  = \frac{1}{n}      \,  ,
\label{mixed}
\end{eqnarray}
where $z_{Q_R}$ is the  $U(1)_B$ charge of $Q_R$, etc.
The $U(1)_Y [U(1)_B]^2$ anomaly then cancels only if
\begin{eqnarray}
z_{Q_L} = 2 z_{U_R} - z_{D_R}~. 
\label{eq:y1B2}
\end{eqnarray}
It follows that there is no $U(1)_B$ gauge-gravitational anomaly, and 
finally, the  $[U(1)_B]^3$ anomaly cancels only if 
\begin{equation}
\left( z_{U_R} - z_{D_R} \right) \left( 7z_{U_R} - z_{D_R} +
3/n \right) = 0 ~.
\end{equation}
We will refer to the $z_{D_R} = z_{U_R}$ solution as the {\it D=U}
model, and to the $z_{D_R} = 7 z_{U_R} + 3/n$ solution as the
{\it D=7U+3} model. Both these models are in fact families of $U(1)_B$
charges for the vectorlike quarks described by a rational parameter
($z_{U_R} \equiv z$) and an integer $n$ (the number of vectorlike
flavors).

\begin{table}[t]
\renewcommand{\arraystretch}{1.2}
\begin{tabular}{c|cc|c}\hline 
field  & $SU(2)_W$   &  $U(1)_Y$  & $U(1)_B$ \ \ \ \\  [-0.03em] \cline{4-4}
 & & &   \hspace*{-1.5mm}   $D\!=\!U\,$model  \hspace*{-1.5mm}   \vline  
         \hspace*{-.5mm}  $D\!=\!7U\!\!+\!3\,$model \\ \hline\hline
$u_R^j$ & \  1 \ & \  $\!\!+2/3$ &  $+1/3$ \ \ \ \\ [0.1em]
$d_R^j$ & \  1 \ & \  $\!\!-1/3$  &   $+1/3$ \ \ \ \\ [0.1em]  
$q_L^j$ & \  2 \ & \  $\!\!+1/6$  &  $+1/3$ \ \ \ \\ [0.1em]
\hline
$U_L^k$ &         &                   &   $z+ 1/n$  \ \ \ \\ [-1.3em]
                 & \  1 \ & \  $\!\!+2/3$  &    \\ [-1.3em]
$U_R^k$ &         &                   &  $z$ \ \ \ \ \\ [0.1em]      
 \hline
$D_L^k$ &          &                   &  \hspace*{2.3mm}  $z+ 1/n$  \hspace*{4.2mm}   \vline   \hspace*{4mm}  \ $7 z + 4/n$ \hspace*{0.1mm}  \\ [-1.3em]
                 & \  1 \ & \  $\!\!-1/3$  &    \\ [-1.3em]
$D_R^k$ &         &                     &  \hspace*{5.5mm}  $z$  \hspace*{11.mm}               \vline  \  \hspace*{4mm}   $7 z + 3/n$ \hspace*{0.1mm}  \\ [0.1em]
  \hline
$Q_L^k$  &         &                   &  \hspace*{8mm}   $z$   \hspace*{11mm}          \vline  \   \hspace*{2mm}  $-5 z - 3/n$ \hspace*{2mm}  \\  [-1.3em]
                  & \  2 \ & \ $\!\!+1/6$  &                                                     \\  [-1.3em]
$Q_R^k$ &         &                   & \hspace*{5.mm}   $z+ 1/n$   \hspace*{4.1mm}   \vline  \   \hspace*{2mm}  $-5 z - 2/n$  \hspace*{2mm} \\ [0.1em]     
\hline
$\phi$ & \  1 \ & \  0 \  &   $+1/n$  \ \ \ \\ \hline
\end{tabular}
\medskip \\
\vspace{0.2in}
\caption{\small Fields carrying $U(1)_B$ charge. With the exception of
  $\phi$ (a color-singlet scalar), all fields shown here are
  color-triplet fermions. The charge assignments labelled by $D\!=\!U$
  and $D\!=\!7U\!+\!3$ correspond to the two solutions of the
  $[U(1)_B]^3$ anomaly cancellation condition.  The SM quarks have a
  generation index, $j = 1,2,3$, and the vectorlike quarks have a
  flavor index $k = 1, \ldots , n$.}
\label{table:U1B}
\end{table}

There is need for at least one scalar field, $\phi$, to carry $U(1)_B$
charge and to have a VEV.  The `vectorlike' quarks are chiral with
respect to $U(1)_B$, so that they can acquire mass only by coupling to
the VEVs that break $U(1)_B$. In renormalizable models with only one
$\phi$ scalar, Eq.~(\ref{mixed}) then requires that the charge of
$\phi$ is $+1/n$ (charge $-1/n$ is the same modulo the interchange of
$\phi$ and $\phi^\dagger$) so that operators of the type $\bar{Q}_L
Q_R \phi^\dagger$ are gauge invariant.  The fields charged under
$U(1)_B$ are listed in Table~\ref{table:U1B}.

If the vectorlike quarks are stable, then they form QCD bound states.
The lightest of these is a heavy-light meson involving a vectorlike
quark and a $u$ or $d$ quark.  If this heavy-light meson is
electrically neutral, then it can be a component of dark matter.  The
heavy-light meson, however, interacts with nucleons ({\it e.g.}, via
meson exchange) and also has a magnetic dipole moment, so that there
are stringent limits on the mass of the vectorlike quark from direct
searches for dark matter.

A simpler alternative is that all vectorlike quarks decay into SM
particles. For that to happen, the charge $z$ must take certain
values, or else there must be additional scalars with VEVs carrying
$U(1)_B$ charge.

In the simplest case, the {\it D=U} model with $n = 1$
\cite{FileviezPerez:2011pt}, we find that $z=1/3$ would allow decays
into a SM quark and $h^0$ (the SM Higgs boson) through the Yukawa
interactions
\begin{equation}
\bar{q}_L U_R  H \; ,  \; \bar{q}_L D_R H^\dagger \; , \;  \bar{u}_R Q_L  H^\dagger \; ,  \; \bar{d}_R Q_L  H ~, 
\end{equation}
where $H$ is the SM Higgs doublet.  In addition, decays into a SM
quark and $\phi$ may proceed through the following Yukawa terms:
\begin{equation}
\bar{q}_L Q_R \phi^\dagger   \; ,  \;    \bar{u}_R U_L \phi^\dagger   \; ,  \;    \bar{d}_R  D_L \phi^\dagger  \; .
\label{decay-operators}
\end{equation}
Even when the two particles described by the complex scalar $\phi$ are
heavier than the vectorlike quarks, the above Yukawa terms in
conjunction with the Higgs portal coupling $H^\dagger H \phi^\dagger
\phi$ induce $Q, U$ and $ D$ decays, of the type $Q \to q h^0 h^0$ (or
$Q \to q\, b \bar{b} \, h^0 $ for masses below $2 M_h$).  For
$z=-5/3$, the vectorlike quarks decay through Yukawa couplings of the
type $\bar{q}_L Q_R \phi$.

The {\it D=7U+3} model with $n = 1$ has different decay patterns. For
example, $z = -2/3$ implies that the decays $Q_R \to q_L \phi$ and
$D_L \to d_R \phi^\dagger$ are allowed, but $U$ can decay via
renormalizable interactions only if there is at least one additional
field ({\it e.g.}, a scalar $S$ which is a SM gauge singlet, has
$U(1)_B$ charge 0, and interacts through $\bar{u}_R U_L S$).

The {\it D=U} and {\it D=7U+3} models are identical for $z =
  -1/(2n)$.  In this case, a second scalar $\phi^\prime$, of
  $U(1)_B$ charge $1/3 -1/(2n) $, is necessary to allow $Q$, $U$, and $D$
  decays through $\bar{q}_L Q_R \phi^\prime$, $\bar{u}_R U_L
  \phi^\prime$ and $\bar{d}_R D_L \phi^\prime$, respectively.  
  
The choice of vectorlike fermions shown in Table~\ref{table:U1B} is
simple but not unique. For example, anomaly cancellation in the
presence of vectorlike leptons instead of quarks is also possible
\cite{Carena:2004xs}.  A fourth generation of chiral quarks and
leptons can also lead to the cancellation of the $U(1)_B$ anomalies
\cite{Carone:1994aa}, but this possibility is nearly ruled out
\cite{Kumar:2012ww} now by the measurements of Higgs production
through gluon fusion \cite{Chatrchyan:2013sfs}, and by direct searches
for $t^\prime$ \cite{tprime} and $b^\prime$
\cite{bprime} quarks at the LHC.

The couplings of the $Z^\prime_B$ to SM quarks are given by
\begin{equation}
\frac{g_B}{6}  Z^\prime_{ B \mu}  \,  \overline{q}\gamma^\mu q \ ,
\end{equation}
where $g_B$ is the $U(1)_B$ gauge coupling (using the normalization
where the group generator is 1/2), and is related to the coupling
constant, as usual, by $\alpha_B = g_B^2/(4\pi)$.  The $Z^\prime_B$
can decay into a pair of jets (including $b$ jets) or into a
$t\bar{t}$ pair (for a $Z^\prime_B$ mass $M_{Z^\prime_B} > 2 m_t$),
with partial decay widths given by
\begin{eqnarray}
\Gamma\!\left(Z^\prime_B\! \to jj\right) &=& \frac{5  \alpha_B}{36}  M_{Z^\prime_B}  \left( 1 + \frac{\alpha_s}{\pi} \right)  ~, \nonumber \\ [-4mm]
\\ [1mm]
\frac{\Gamma\!\left(Z^\prime_B\! \to t\bar{t}\right)}{\Gamma\!\left(Z^\prime_B\! \to jj\right)} &=&  \frac{1}{5} \!\left( \! 1 \!-\!  \frac{4 m_t^2}{M_{Z^\prime_B}^2 } \!\right)^{\!\!1/2} \!
\left[ 1\! + O\!\left(\frac{\alpha_s m_t}{M_{Z^\prime_B}} \right) \! \right] .
\nonumber
\end{eqnarray}
Here we have included the NLO QCD corrections and no electroweak
corrections.  If the decays into vectorlike quarks are kinematically
closed, then the total width of $Z^\prime_B$ is
\begin{equation}
\Gamma_{Z^\prime_B}  =  \Gamma\!\left(Z^\prime_B \to jj\right)  +  
\Gamma\!\left(Z^\prime_B \to t\bar{t}\right) ~~.
\end{equation}

\subsection{Coloron}
\label{subsec:coloron}

Another hypothetical particle that can easily produce dijet resonances
with large cross section at the LHC is the coloron \cite{Hill:1991at},
a spin-1 color-octet gauge boson. The coloron, in the case of
flavor-universal couplings \cite{Chivukula:1996yr}, is not
significantly constrained by flavor processes nor by other low energy
data.  Furthermore, the coloron is automatically leptophobic.

The simplest gauge symmetry that can be associated with a heavy
color-octet vector boson is $SU(3)_1 \times SU(3)_2$
\cite{Hall:1985wz}. This is spontaneously broken down to the diagonal
$SU(3)_c$ gauge group, which is identified with the QCD one. A minimal
renormalizable extension of the SM which includes a coloron, dubbed
ReCoM, is analyzed in Ref.~\cite{Bai:2010dj}. Assuming that all the SM
quarks transform as $(3,1)$ under $SU(3)_1 \times SU(3)_2$, the
couplings of the coloron to SM quarks are given by the Lagrangian term
\begin{equation}
g_s \tan{\theta}\;\overline{q}\gamma^\mu T^a G_\mu^{\prime\,a}q  ~~,
\end{equation}
where $g_s = \sqrt{4 \pi \alpha_s}$ is the QCD gauge coupling and
$\tan{\theta} > 0$ is a dimensionless parameter.

If there are no new quarks mixing with the SM ones, and no additional
color-octet spin-1 particles, then $\tan{\theta}$ is the ratio of the
$SU(3)_2$ and $SU(3)_1$ gauge couplings.  These gauge couplings can
vary between $g_s$ and some upper limit of about $4\pi /\sqrt{3}$
corresponding to the nonperturbative regime. Consequently, there are
both upper and lower limits on $\tan{\theta}$ \cite{Dobrescu:2009vz}:
$0.15 \lesssim \tan{\theta} \lesssim 6.7$.  Unlike the $Z^\prime_B$,
whose UV behavior requires some new fermions, the flavor-universal
coloron is anomaly free. Nevertheless, vectorlike quarks may be
present, and if they mix with the SM quarks, then the lower limit on
$\tan{\theta}$ no longer applies \cite{Dobrescu:2007yp}. Similarly, a
second heavy spin-1 color-octet particle can mix with the coloron and
dilute its couplings to quarks.

The partial decay widths of the coloron of mass $M_{G^\prime}$ into
$jj$ (including $b\bar{b}$) and into $t\bar{t}$ are given by
\begin{eqnarray}
\label{coloron-width}
\Gamma\!\left(G^\prime \! \to jj\right) &=& \frac{5 \alpha_s}{6}  \tan^2{\!\theta}\; M_{G^\prime}  \left[ 1 \!+ O\left(\frac{\alpha_s}{\pi} \right)\! \right]   ~, \nonumber \\ [-2mm]
\\ [-1mm]
\frac{\Gamma\!\left(G^\prime \! \to t\bar{t}\right)}{\Gamma\!\left(G^\prime\! \to jj\right)} &=&  \frac{1}{5} \!\left( \! 1 \!-\!  \frac{4 m_t^2}{M_{G^\prime}^2 } \right)^{\!\!1/2} \!
 \left[ 1\!+ O\!\left(\frac{\alpha_s m_t}{M_{G^\prime}} \right)\! \right]  \ ,
\nonumber
\end{eqnarray}
where only NLO QCD corrections are included.

The minimal scalar sector responsible for breaking the $SU(3)_1 \times
SU(3)_2$ symmetry (which is part of ReCoM) includes a color octet and
two color singlets.  If these are light enough, then the coloron can
decay into two octet scalars or into an octet scalar and a singlet
scalar, with partial decay widths that are especially large for
$\tan{\theta} \ll 1$ \cite{Bai:2010dj}. In what follows, we will
assume that the scalars are heavier than $M_{G^\prime}/2$, so that the
total width of the coloron is simply the sum of the $jj$ and
$t\bar{t}$ partial widths shown in Eq.~(\ref{coloron-width}).

\section{Collider searches of dijet resonances}
\label{sec:Searches}

We now detail our procedure and results for mapping the existing dijet
resonance searches to the coupling--mass plane.

\subsection{Mapping procedure and experimental limits}
\label{subsec:procedure}

As discussed in Section I, the partons responsible for $s$-channel
production at hadron colliders are also a decay mode, and so the new
particle must decay back to pairs of jets at some rate.  Models that
give rise to a spin-1 dijet resonance are the most straightforward to
construct.  For the representative spin-1 particles discussed in
Section~\ref{sec:Models}, the $Z^\prime_B$ boson and the coloron,
there are two parameters that characterize the dijet signal: mass and
coupling.

Even with only two parameters, the extraction of limits from
experimental searches for dijet resonances remains challenging.  For
example, varying the resonance mass while keeping the coupling fixed
introduces varying levels of final state radiation, cut-dependent
effects from parton distribution function (PDF) sampling at high
masses relative to the total $\sqrt{s}$ (the mass dependence of PDFs
is shown in \cite{Quigg:2009gg}), and trigger-dependent efficiencies
at low masses.

Dijet resonance searches probe the existence of narrow peaks in the
dijet invariant mass ($m_{jj}$) spectrum.  The QCD background is
expected to be a smoothly falling exponential.  Other backgrounds,
such as hadronic $t \bar{t}$ decays, are expected to give broad
features at their respective mass scales.

Although a bump-like feature on top of a smoothly falling background
is seemingly easy to observe, the experimental resolution in the dijet
channel is rather poor ($\sim 5-10$\%, depending on
mass~\cite{Harris:2011bh} as well as experiment), and the QCD
background at energies much smaller than the total $\sqrt{s}$ become
overwhelming.  Higher $\sqrt{s}$ colliders rapidly lose sensitivity to
low mass resonances in dijet searches because of the minimum $p_T$,
$E_T$, and $m_{jj}$ trigger requirements.  Pre-scaled triggers (and
so-called ``data-scouting'' techniques~\cite{CMS-PAS-EXO-11-094,
  CMS:2012ooa}), however, can help augment the trigger bandwidth to
extend the searches down to lower masses.

For our mapping, we start by running a Monte Carlo (MC) simulation for
a given choice of coupling and mass.  In the narrow width
approximation, the $s$-channel production factorizes from the decay,
hence the acceptance and efficiency do not depend on the coupling at
leading order. Some dependence on the coupling arises from loops
involving the new spin-1 particle, as shown in the case of NLO coloron
production \cite{Chivukula:2011ng}; however, this effect is relatively
small, and for simplicity we ignore it in what follows.  For a given
set of experimental cuts we obtain a simulated effective rate. The
ratio of the experimental limit to the simulated effective rate is the
square of a coupling rescaling factor. Multiplying the initial
coupling by this rescaling factor gives the experimental upper limit
on the coupling for the chosen mass.  Repeating this procedure for all
experimental searches, we determine the excluded region in the
coupling--mass plane.

There has been a host of resonance searches from every experiment at
hadron colliders in the dijet channel.  We summarize all of them in
Table~\ref{table:expts}.

\begin{table}
\begin{tabular}{c|c|c|c| rcl}\hline
Collisions,& Experiment & Ref. &  Luminosity & \multicolumn{3}{|c}{ Mass range } \\
$\! \sqrt{s}$ (TeV) &  & &  (fb$^{-1}$)  & \multicolumn{3}{|c}{(TeV)} \\
\hline
\hline
                  &  \parbox{1cm}{~ \\ [1mm] UA2 \\ [-1mm]}   & \cite{Alitti:1990kw} & $4.7 \times 10^{-3}$ & \  0.08 &--& 0.32 \\
$p \bar{p}$, 0.63 &  & \cite{Alitti:1993pn} & $1.1 \times 10^{-2}$ & 0.14 &--& 0.3 \\
\cline{2-7}
                  & UA1 & \cite{Albajar:1988rs} & $4.9 \times 10^{-4}$ &  0.15 &--& 0.4 \\  

\hline
                 &  & \cite{Abe:1989gz} & $2.6 \times 10^{-6}$ & 0.06 &--& 0.5 \\
                 &  \parbox{1cm}{~ \\ [1mm] CDF \\ [-1mm]}  & \cite{Abe:1993it} & $4.2 \times 10^{-3}$ & 0.2 &--& 0.9 \\
$p \bar{p}$, 1.8 &  & \cite{Abe:1995jz} & $1.9 \times 10^{-2}$ & 0.2 &--& 1.15 \\
                 & & \cite{Abe:1997hm} & $0.11$ & 0.2 &--& 1.15 \\
\cline{2-7}
                 & D0 & \cite{Abazov:2003tj} & $0.11$ & 0.2 &--& 0.9 \\
\hline
$p \bar{p}$, 1.96 & CDF & \cite{Aaltonen:2008dn} & $1.1 \; $ & 0.26 &--& 1.4 \\
\hline
        &  & \cite{Aad:2010ae} & $3.2 \times 10^{-4}$ & 0.3 &--& 1.7 \\
        & & \cite{Aad:2011aj}  & $3.6 \times 10^{-2}$ & 0.6 &--& 4 \\
        &  \parbox{1.6cm}{~ \\ [1mm]  ATLAS \\ [-1mm]}  & \cite{ATLAS-CONF-2011-081} & $\, 0.16$ & 0.9 &--& 4 \\
        &  & \cite{ATLAS-CONF-2011-095} & $\, 0.81$ & 0.9 &--& 4 \\
$pp$, 7 & & \cite{Aad:2011fq} & $1.0 \; $ & 0.9 &--& 4 \\
        &  & \cite{ATLAS:2012pu} & $4.8 \; $ & 1 &--& 4 \\
\cline{2-7}
        & & \cite{Khachatryan:2010jd} & $2.9 \times 10^{-3}$ & 0.5 &--& 2.6 \\
        &  \parbox{1cm}{~ \\ [1mm] CMS \\ [-1mm]}  & \cite{Chatrchyan:2011ns} & $1.0 \; $ & 1 &--& 4.1 \\
        & & \cite{CMS:2012yf} & $5.0 \; $ & 1 &--& 4.3 \\
        &  & \cite{CMS-PAS-EXO-11-094} & $\, 0.13$ & 0.6 &--& 1 \\
\hline
        &  \parbox{1.6cm}{~ \\ [1mm]  ATLAS \\ [-1mm]}  & \cite{ATLAS-CONF-2012-088} & $5.8 \; $ & 1.5 &--& 4 \\
$pp$, 8  &  & \cite{ATLAS-CONF-2012-148} & $13 \;\;\;\; $ & 1.5 &--& 4.8 \\
\cline{2-7}
        &  \parbox{1cm}{~ \\ [1mm]  CMS \\ [-1mm]}  & \cite{Chatrchyan:2013qha} & $4.0 \; $ & 1 &--& 4.8 \\
        &  & \cite{CMS-PAS-EXO-12-059} & $20 \;\;\;\; $ & 1.2 &--& 5.3 \\
\hline
\end{tabular}
\caption{Mass ranges for existing dijet resonance searches at hadron colliders.}
\label{table:expts}
\end{table}

For each mass point and collider, we simulate an event sample of
on-shell $s$-channel $Z^\prime_B$ as well as coloron production (at
leading order) with subsequent decay to light-flavor and $b$ jets
using MadGraph 5 v1.5.7~\cite{Alwall:2011uj} with the CTEQ6L1
PDFs~\cite{Pumplin:2002vw}.  Each event is passed through
\textsc{Pythia} v6.4.20~\cite{Sjostrand:2006za} for showering and
hadronization, and then through PGS v4~\cite{PGS4} for basic detector
simulation and jet clustering.  

Choosing $g_B =0.2$ for the $Z^\prime_B$ or $\tan \theta = 0.2$ for the coloron, 
we obtain the cross section times branching fraction from MadGraph 5,
denoted $\sigma_{0.2} \timesdot \brf$, as a function of mass for each
collider.  We then implement the various triggers and cuts as
described in each analysis listed in Table~\ref{table:expts}, to
obtain an acceptance $A$.  The ratio of the resulting simulated
effective rate, $\sigma_{0.2} \timesdot \brf \timesdot A$, to the
limit from each relevant analysis in Table~\ref{table:expts} allows us
to extract the upper limit on the coupling as a function of mass:
\begin{equation}
(g_B)_{\rm max} = 0.2 \left( \dfrac{ \left( \sigma \timesdot \brf
  \timesdot  A \right)_{\text{limit}} }{ \sigma_{0.2} \timesdot  \brf
  \timesdot  A } \right)^{\! 1/2} ~~,
\end{equation}
and similarly for $(\tan\theta)_{\rm max}$.

We now discuss the most relevant searches, grouped according to the
mass range probed.

\subsubsection{Searches for $m_{jj} < 200$ GeV}
 
Although a couple of searches (from UA2 and CDF, see Table II) extend
to masses below 140 GeV, we do not use them because those results were
based on a subtraction of the expected $W$ and $Z$ dijet distribution
calculated at $O(\alpha_s)$; modern precision of $W$ and $Z$ two-dijet
distributions far exceeds the interpretative power of the effective
rate limit in the $60-140$ GeV mass window.  For studies of
theoretical constraints on colorons in that mass range, see
Ref.~\cite{Krnjaic:2011ub}.

In the $140 - 200$ GeV mass range, by far the largest data sample
(10.9 pb$^{-1}$) has been analyzed~\cite{Alitti:1993pn} by the UA2
experiment at the CERN SPS collider, which operated mainly at
$\sqrt{s} = 630$ GeV.  Mapping the UA2 limit to the coupling--mass
plane is simpler than the procedure required for later analyses
because the UA2 analysis includes a table of efficiencies for
selecting the signal events from background as well as for isolating
the peak feature in the signal events (cf.~Table 1 and Table 2 of
Ref.~\cite{Alitti:1993pn}).  We linearly interpolate this overall
efficiency to obtain a combined acceptance times efficiency factor.
Finally, since the UA2 constraint is presented as a branching fraction
limit on a sequential SM $Z'$, we unfold the $Z'$ cross section to
obtain an estimated $\sigma \timesdot \brf$ limit, as discussed in
Ref.~\cite{Yu:2011cw}.  After applying the overall efficiency, we
obtain an effective rate limit from UA2, which we then map into the
upper limit on the $Z^\prime_B$ coupling shown on the left-hand side
of Figure \ref{fig:Zplimits}.

\subsubsection{Searches in the 200 -- 900 GeV mass range}

The CDF~\cite{Abe:1997hm} and D0~\cite{Abazov:2003tj} searches using
the full data samples ($\approx 110$ pb$^{-1}$) of the Run I at the
Tevatron compete for the best limit in the $200-260$ GeV mass window.
We choose to extract a limit from the CDF analysis, because it applies
to a larger mass range.

Above $260$ GeV, the CDF analysis using 1.1 fb$^{-1}$ of Run II
data~\cite{Aaltonen:2008dn} supersedes the Run I results.  For the
$260-900$ GeV window, the only ATLAS~\cite{Aad:2010ae, Aad:2011aj} and
CMS~\cite{Khachatryan:2010jd, CMS-PAS-EXO-11-094} searches use very
small $\sqrt{s} = 7$ TeV data sets.  The most competitive of these is
the 0.13 fb$^{-1}$ CMS~\cite{CMS-PAS-EXO-11-094} analysis via
so-called ``data scouting'', which uses a reduced data format to
record events with sensitivity at low masses even during high
instantaneous luminosity conditions.  We implement the appropriate
trigger and analysis requirements for each mass point probed by these
CDF and CMS analyses to calculate their respective acceptances needed
for the coupling--mass mapping.  For all CMS limits, we adopt their
$qq$ resonance constraint, since our spin-1 resonances only couple (at
leading order) to $q \bar{q}$ (note that the radiation patterns from
$qq$ and $q \bar{q}$ final states are indistinguishable).  As can be
seen from Figure \ref{fig:Zplimits}, the CDF limits on spin-1
resonances are the most stringent ones, even though they are based on
only a tenth of the Run II data.

\subsubsection{Searches for resonance masses above 900 GeV}

Most ATLAS and CMS searches begin at about 900 GeV.  For the
$900-1000$ GeV range, the ATLAS 1 fb$^{-1}$ search~\cite{Aad:2011fq}
is expected to be the most sensitive, as it has higher energy than CDF
1.1 fb$^{-1}$~\cite{Aaltonen:2008dn}, and a larger data sample than
the other ATLAS~\cite{Aad:2010ae, Aad:2011aj, ATLAS-CONF-2011-081,
  ATLAS-CONF-2011-095} and CMS studies~\cite{Khachatryan:2010jd,
  CMS-PAS-EXO-11-094}.

From $1-1.2$ TeV, the CMS 4.0 fb$^{-1}$ search using 8 TeV
data~\cite{Chatrchyan:2013qha} is expected to be competitive with the
earlier 7 TeV ATLAS 4.8 fb$^{-1}$~\cite{ATLAS:2012pu} and CMS 5.0
fb$^{-1}$ analyses~\cite{CMS:2012yf}, superseding the ATLAS and CMS 1
fb$^{-1}$ analyses~\cite{Aad:2011fq, Chatrchyan:2011ns}.  The slightly
smaller amount of integrated luminosity analyzed in
Ref.~\cite{Chatrchyan:2013qha} compared to Refs.~\cite{ATLAS:2012pu,
  CMS:2012yf} is counterbalanced by the slight increase in collider
energy, giving comparable coupling sensitivities.

\begin{figure*}[t]
\begin{center}
\includegraphics[width=0.83\textwidth, angle=0]{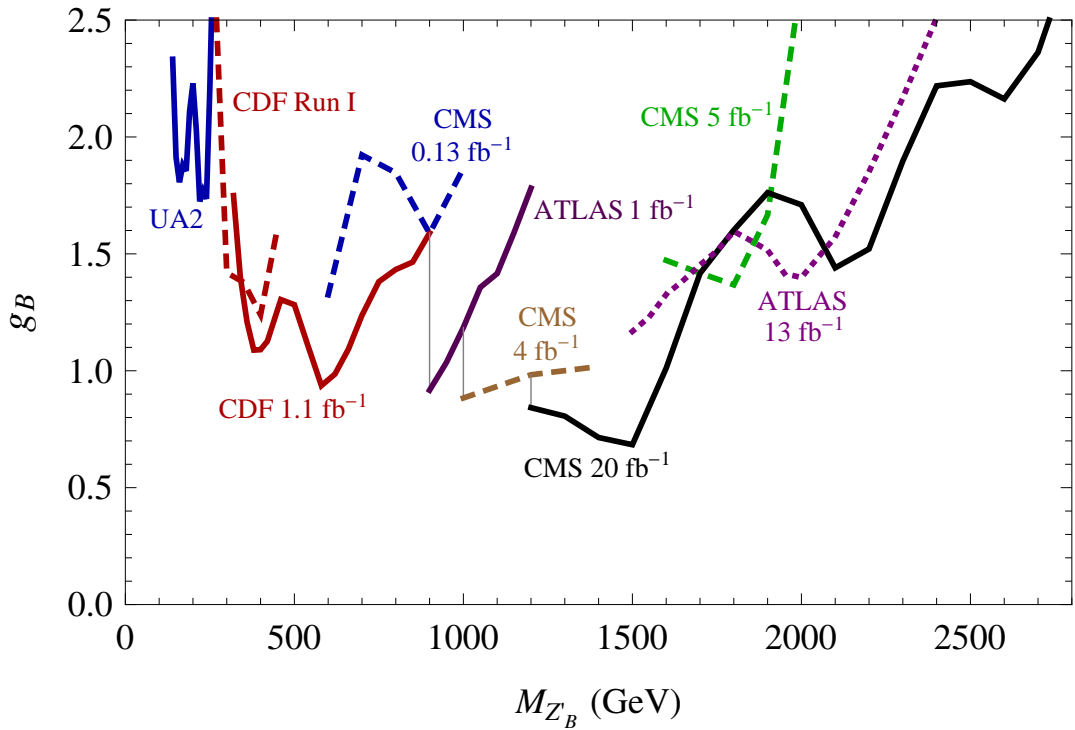}
\caption{Leading experimental limits in the coupling $g_B$ versus mass
  $M_{Z^\prime_B}$ plane for $Z^\prime_B$ resonances. Values of $g_B$ above each
  line are excluded at the 95\% C.L.}
\label{fig:Zplimits}
\end{center}
\end{figure*}

Above $1.2$ TeV, the CMS analysis~\cite{CMS-PAS-EXO-12-059} using 19.6
fb$^{-1}$ of $\sqrt{s} = 8$ TeV data is expected to be the most
sensitive (we will refer to this search as CMS 20 fb$^{-1}$).  This
analysis is the most recent dijet resonance search and has sensitivity
to resonances as heavy as 5 TeV.  Nevertheless, upward fluctuations in
the CMS 20 fb$^{-1}$ limit actually leave some small gaps where the
ATLAS~\cite{ATLAS-CONF-2012-148} $13.0$ fb$^{-1}$ limit and the CMS
$5.0$ fb$^{-1}$ limits are more stringent (see Figures
\ref{fig:Zplimits} and \ref{fig:Colimits}).

For the various CMS analyses, we implement the respective trigger and
analysis cuts to tabulate the acceptance for each mass point and
obtain an effective rate limit from the $qq$ resonance constraint. For
the 8 TeV analyses we find that the acceptance grows from 38\% to 50\%
for the $Z'_B$ signal (and from 33\% to 44\% for the coloron) when the
mass grows from 1 TeV to 2.5 TeV, and is constant at larger masses.

The ATLAS results, on the other hand, are presented as limits on
Gaussian resonances in the $m_{jj}$ spectrum after trigger
requirements, detector effects, and analysis cuts are implemented.
This poses additional problems because the $m_{jj}$ distribution
produced by any particle decaying to a pair of jets would be quite
different from a Gaussian: a long tail at low invariant mass is
induced by imperfect recapturing of final state radiation.  To
overcome this mismatch, we form the $m_{jj}$ spectrum after cuts and
then model and apply a Gaussian core efficiency ($\epsilon_G$) for our
signal, which is an additional factor beyond the canonical
acceptance. Our procedure of determining $\epsilon_G$ is described in
the Appendix.

The ATLAS limits are given for a variety of Gaussian widths: for each
ATLAS limit, we adopt the smallest Gaussian width constraint in
performing our mapping, since our resonances typically have intrinsic
widths at the percent and sub-percent level.

\subsection{Results and Discussion}

Following the procedure described in Section~\ref{subsec:procedure}
for the leading experimental dijet limits, we obtain the
coupling--mass mapping shown in Figure~\ref{fig:Zplimits} for a spin-1
resonance $Z^\prime_B$. We have used the leading order production
computed with MadGraph 5.

If the NLO corrections to the process $pp \to Z^\prime_B X \to jj X$
are included in an event generator, then the mapping can be performed
more precisely. We do not expect that they would change the values of
$g_B$ by more than $O(10\%)$.

We emphasize that, unlike the usual $\sigma \timesdot \brf \timesdot
A$ limit plots, Figure~\ref{fig:Zplimits} readily shows the exclusion
in coupling as well as mass.  This mapping also demonstrates the
complicated interplay between different experimental analyses using
different colliders and luminosities.

From Figure~\ref{fig:Zplimits}, we conclude that $Z^\prime_B$ bosons
are unconstrained for a gauge coupling $g_B \lesssim 0.6$, leaving a
large area of parameter space unexplored by dijet resonance searches
so far.  Moreover, for the entire sub-TeV region, the experimental
limits allow for $g_B$ couplings as large as 0.9, while locally, $g_B$
couplings can reach $\approx 1.5$.  Importantly, an update from CDF or
an analysis by D0 with their full $\approx 10$ fb$^{-1}$ Run II
datasets could offer evidence for or provide interesting limits on new
sub-TeV dijet resonances.  We also note that an update of the ``scouted
data'' analysis \cite{CMS-PAS-EXO-11-094} with more luminosity by CMS
(and ATLAS) would also push sensitivity to lower couplings in the
several hundred GeV mass range.

The plot is not extended above $g_B = 2.5$, because the $U(1)_B$
coupling constant is already large, $\alpha_B = g_B^2/(4\pi) \approx
0.5$, so that it is difficult to avoid a Landau pole.  For that large
coupling, the current mass reach is around 2.8 TeV. The 14 TeV LHC
will extend significantly the mass reach, and can probe smaller
couplings once enough data is analyzed.  Note that couplings of $g_B
\approx 0.1$ can be viewed as typical (the analogous coupling of the
photon is approximately 0.3), and even $g_B$ as small as 0.01 would
not be very surprising.

\begin{figure*}[t]
\begin{center}
\includegraphics[width=0.83\textwidth, angle=0]{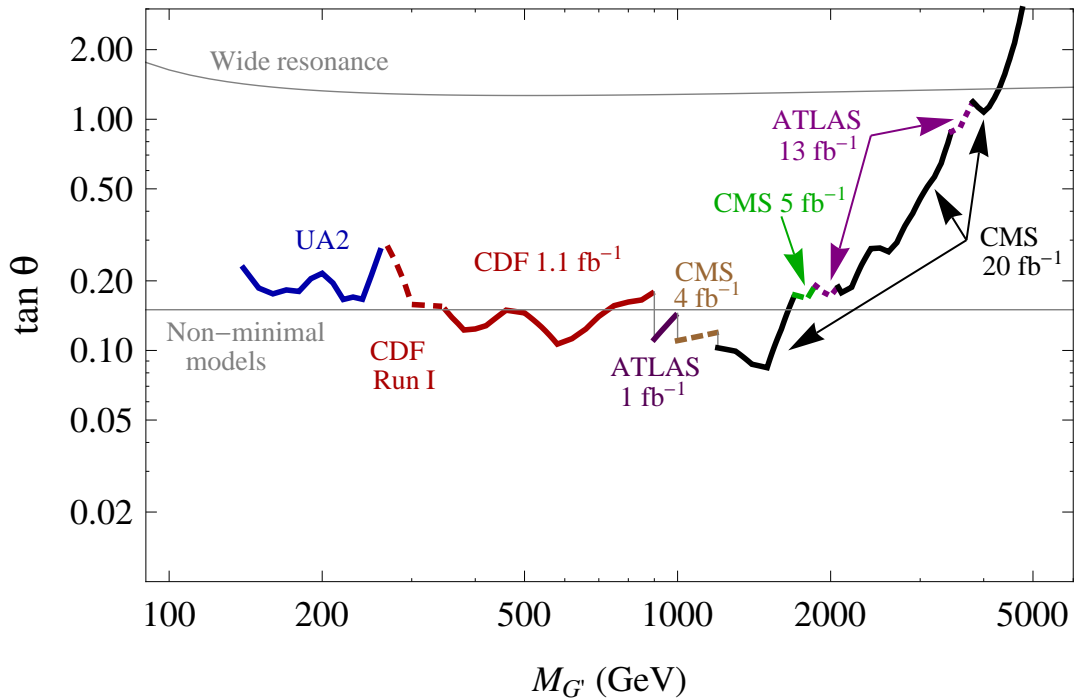}
\caption{Envelope of experimental limits for colorons of coupling
  $\tan \theta$ and mass $M_{G^\prime}$.  Values of $\tan \theta$
  above each limit are excluded at the 95\% C.L.  The region above the
  line labelled ``Wide resonance" corresponds to a coloron
  width-to-mass ratio larger than 15\%. The region below the line
  labelled ``Non-minimal models" (at $\tan \theta = 0.15$) requires
  additional particles, such as vectorlike quarks or a second coloron.
}
\label{fig:Colimits}
\end{center}
\end{figure*}

We also present the coupling--mass mapping for colorons in
Figure~\ref{fig:Colimits}.  For clarity, we only show the envelope of
the strongest $\tan \theta$ upper limits from all available analyses
at each coloron mass.  This mapping is performed again using leading
order production. The NLO corrections to coloron production have been
computed recently \cite{Chivukula:2011ng}, and can vary between
roughly $-30\%$ and $+20\%$.  We do not take the NLO corrections into
account as we do not have an event generator that includes them;
furthermore, there is some model dependence in the NLO corrections at
small $\tan \theta$ (for example, they are sensitive to the
color-octet scalar present in ReCoM~\cite{Bai:2010dj}).

As mentioned in Section~\ref{subsec:coloron}, in the minimal coloron
model there is a theoretical lower limit imposed by perturbativity,
$\tan \theta \gtrsim 0.15$.  The experimental upper limit dips below
that value only in the $350 - 700$ GeV and $0.9-1.6$ TeV mass ranges.
In non-minimal models, where there are vectorlike quarks or a second
coloron, $\tan \theta$ can be substantially smaller than 0.15. Thus,
searches for colorons should continue even after they rule out $\tan
\theta \gtrsim 0.15$ in some mass windows. On the contrary, the
discovery of a coloron with $\tan \theta < 0.15$ would imply the
existence of additional colored particles that can be probed in hadron
collisions.

Unlike the $U(1)$ gauge bosons, the coloron can be rather strongly
coupled before reaching the perturbative upper limit, $\tan \theta
\lesssim 6.7$, because it is associated with a non-Abelian gauge
interaction that is asymptotically free.

There is, however, a tighter upper limit on $\tan \theta$ if the total
width, $\Gamma\!\left(G^\prime \to jj\right) + \Gamma\!\left(G^\prime
\to t\bar{t}\right)$ [see Eq.~(\ref{coloron-width})], is required to
be smaller than the dijet resolution.  A ratio of the total width to
$M_{G^\prime}$ of 15\% (as used in \cite{Aad:2011fq}) corresponds to
$\tan \theta \approx 1.3$ for $M_{G^\prime} \gg 2 m_t$ (with a mild
dependence on $M_{G^\prime}$ due to the running of $\alpha_s$), and to
slightly larger $\tan \theta$ for smaller masses, as shown in
Figure~\ref{fig:Colimits}.  Limits above the line marked there by
``Wide resonance" are not reliable if set by narrow resonance searches
\cite{Choudhury:2011cg}.  Clearly, resonances that have a much broader
intrinsic width than the experimental $m_{jj}$ resolution will fade
more easily into the exponentially falling QCD background.
Separately, for large enough coupling, the $t$-channel exchange of a
coloron starts contributing significantly to the dijet signal, further
diluting the $m_{jj}$ peak.  Note also that at resonance masses
approaching the total $\sqrt{s}$ of the collider, PDF uncertainties
increase.

\section{Conclusions}
\label{sec:Conclusion}

We urge the experimental collaborations to present limits (or contours
if a signal is observed) on dijet resonances in the
coupling-versus-mass plane of a ``baryonic" $Z^\prime_B$, as in
Figure~\ref{fig:Zplimits} (or a coloron as in
Figure~\ref{fig:Colimits}, if the search is sensitive primarily to
large signals arising from heavy resonances).  This coupling--mass
mapping, while being somewhat model dependent (and thus a complement
to---not a replacement for---the cross-section limit plots), has
multiple advantages.  First, it allows a comparison of limits set by
experiments performed at different colliders, and at different
center-of-mass energies.  Second, it allows an assessment of how
stringent the experimental limits are, by comparing them with the
expected range of the physical coupling.  Third, it provides a direct
interpretation, without the need for MC simulations to compute the
acceptance or to convert limits on a Gaussian into limits on a
realistic particle (in the case of the existing ATLAS results).

The coupling--mass mapping also highlights gaps in the combined
sensitivity of all existing searches.  Figure~\ref{fig:Zplimits} shows
that the coupling reach is rather poor in the mass range of 700--900
GeV, and it is even worse for masses below about 300 GeV.  A new
analysis by CDF or D0 with the full Run II data set could have great
impact there. Non-conventional methods, such as analyzing scouted data
\cite{CMS-PAS-EXO-11-094}, are also important for extending the
sensitivity of LHC experiments in the sub-TeV mass range. More
generally, the traditional trend for each new dijet search to attain
ever higher mass reach does not need to leave the (equally important)
small-coupling region unexplored.

We have argued that the simplest origin of narrow dijet resonances is
a spin-1 particle with flavor-independent couplings.  Our mapping thus
focused on the $Z^\prime_B$ and $G^\prime$.  The same procedure can
also be applied to other spins or color representations
\cite{Han:2010rf}, but the results would be different because of PDF
dependencies and radiation patterns of the decay products.

The overview of theoretical and experimental status of dijet
resonances included in this paper is not exhaustive.  For example, we
have not discussed angular correlations, which complement the
information contained in the $m_{jj}$ distribution.  We also note that
any particle that produces a dijet resonance can also be produced in
association with a $W$, a $Z$ or a photon.  Even though the cross
sections for these associated productions are much smaller
\cite{Bai:2010dj, Atre:2012gj}, the searches for $W + jj$, $Z + jj$,
and $\gamma + jj$ benefit from better triggers that extend sensitivity
to lower resonance masses compared to the pure dijet resonance
searches.
 
The coupling--mass plane can and should be used for any resonance
search (as it has been done in some cases, {\it e.g.},
\cite{Abe:1995jz, Abe:1997hm, Aaltonen:2009qu, Zerwekh:2009vi}).  In
particular, the $t \bar{t}$ resonance searches can be interpreted in
terms of the same $Z^\prime_B$ or coloron. For these flavor-blind
particles, it would also be interesting to investigate the
complementarity between $t \bar{t}$ and dijet resonance searches.

If a dijet resonance will be discovered in the absence of a dilepton
resonance at the same mass, it is likely that additional colored
particles will remain to be discovered. To see this, recall (from
Section \ref{sec:Models}) that the $Z^\prime_B$ requires some
vectorlike fermions to cancel the gauge anomalies, while the coloron
requires at least some scalars from the gauge symmetry breaking
sector.

\section*{Acknowledgements}
\label{sec:Acknowledgements}

We would like to thank Sekhar Chivukula, Arsham Farzinnia, Robert
Harris, Olivier Mattelaer, Elizabeth Simmons and Ciaran Williams for
useful communications.  We are grateful to James Bourbeau for pointing
out an error in an earlier version of Eq.~(3).  Fermilab is operated
by the Fermi Research Alliance, LLC under Contract
No. De-AC02-07CH11359 with the United States Department of Energy.

\section*{Appendix: From Gaussians to particles}

As explained in Section~\ref{subsec:procedure}, the effective rate in
the case of ATLAS analyses is given by $\sigma \timesdot \brf
\timesdot A \timesdot \epsilon_G$, where $\epsilon_G$ is the
efficiency of converting the limits on a realistic particle (whose
$m_{jj}$ distribution has a long tail due to final state radiation)
into limits on a Gaussian. As noted in the Appendix of
Ref.~\cite{Aad:2011fq}, the low-$m_{jj}$ tail should be removed as it
does not contribute to the assumed Gaussian signal.  In this Appendix
we present a more precise procedure for estimating $\epsilon_G$.

We fit the $m_{jj}$ signal spectrum with a Crystal Ball function
\cite{Gaiser:1982yw},
\begin{equation}
\begin{array}{l}
f(x; \alpha, n, \bar{x}, \sigma ) = N \\
 \times \left\{
\begin{array}{c}
\exp \left( - \dfrac{(x - \bar{x})^2}{2 \sigma^2} \right) \ , 
\text{ for } \dfrac{x - \bar{x}}{\sigma} > -\alpha \ , \\
A \left(\! \dfrac{n}{|\alpha|} \!-\! |\alpha| \!-\! 
\dfrac{x - \bar{x}}{\sigma} \right)^{\!\!-n} \!\! ,
\text{ for } \dfrac{x - \bar{x}}{\sigma} \leq -\alpha \ , \\
\end{array} \right.
\end{array}
\label{eqn:crystalball}
\end{equation}
which is a combination of a truncated Gaussian and a power law tail;
here
\begin{equation}
A = \left( \frac{n}{|\alpha|} \right)^{\! n}  e^{-|\alpha|^2/2}  ~~ ,
\end{equation}
and $N$ is an overall normalization factor.  The fit parameters
$\alpha$, $n$, $\bar{x}$, and $\sigma$ correspond to the location of
the power law--Gaussian crossover in units of $\sigma$, the power law
exponent, and the mean and width of the Gaussian, respectively.
Performing this fit allows us to use the Gaussian fit parameters to
calculate the Gaussian core efficiency.

\begin{figure}[t]
\begin{center}
\includegraphics[width=0.43\textwidth, angle=0]{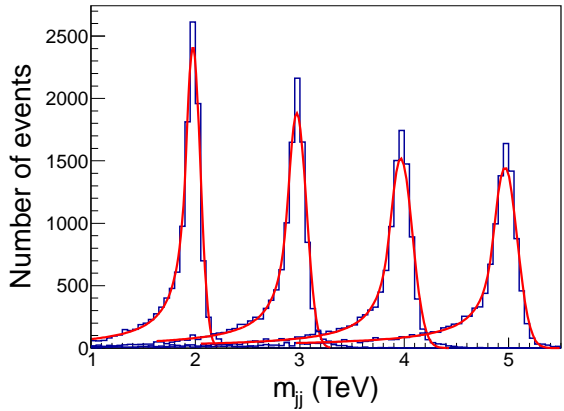}
\caption{Simulated $m_{jj}$ distributions (blue histograms) and
  Crystal Ball fits (red curves) performed on 25000 $Z'$ signal events
  for $M_{Z^\prime_B} = 2$, 3, 4, and 5 TeV, after implementing the CMS 20
  fb$^{-1}$ search~\cite{CMS-PAS-EXO-12-059}.}
\label{fig:crystalball}
\end{center}
\end{figure}

We have checked that this fitting function accurately reproduces the
expected signal shape for our on-shell $s$-channel resonance
production in the $m_{jj}$ spectrum, as shown in
Figure~\ref{fig:crystalball}, which is in reasonable agreement with
Fig.~2 of Ref.~\cite{CMS-PAS-EXO-12-059}\footnote{Even though we only
  use the Gaussian fit procedure on ATLAS limits, we show the fit
  results from our implementation of the CMS 20 fb$^{-1}$ study to
  facilitate comparison with Fig.~2 of Ref.~\cite{CMS-PAS-EXO-12-059}.
  The corresponding $m_{jj}$ spectrum and Crystal Ball fits for ATLAS
  analyses are qualitatively similar.}.  The fit parameters are varied
within a large range of values: $\alpha$ floats within 0.05 to 3.5,
$n$ within 0.5 to 5.0, $\bar{x}$ within (70--105)\% of the truth boson
mass, $\sigma$ within (0.5--30)\% of the truth boson mass, and the
overall normalization $N$ floats within (50--300)\% of the
$N_{\text{events}}^{\text{total}}$ in the histogram.  Moreover, the
beginning and end of the fit range is adjusted from (50--70)\% and
(110--115)\% of the $m_{jj}$ peak, respectively.  The fit range with
the smallest $\chi^2$/ndf dictates the fit parameters $\bar{x}$ and
$\sigma$ used in the estimation of the Gaussian core component of the
$m_{jj}$ shape.

From the fit parameters, we count the number of events in the $m_{jj}$
distribution within $\bar{x} \pm 3 \sigma$ to give the efficiency
$\epsilon_G$ for extracting the Gaussian peak appropriate for the
Gaussian template limits set by ATLAS,
\begin{equation}
\epsilon_G = 
\frac{N_{\text{events}}^{\bar{x} \pm 3 \sigma} }{ N_{\text{events}}^{\text{total }} } \ .
\end{equation}

From our simulated ATLAS $m_{jj}$ distributions, we get $\epsilon_G$
values for both the $Z'_B$ and the coloron between 55\% and 65\%.



\begin{thebibliography}{99}

\bibitem{Han:2010rf} 
  T.~Han, I.~Lewis and Z.~Liu,
  ``Colored resonant signals at the LHC: largest rate and simplest topology,''
  JHEP {\bf 1012}, 085 (2010)
  [arXiv:1010.4309].

\bibitem{Harris:2011bh} 
  R.~M.~Harris and K.~Kousouris,
  ``Searches for Dijet Resonances at Hadron Colliders,''
  Int.\ J.\ Mod.\ Phys.\ A {\bf 26}, 5005 (2011)
  [arXiv:1110.5302]. 

\bibitem{Alitti:1990kw} 
  J.~Alitti {\it et al.}  [UA2 Collaboration],
  ``A measurement of two jet decays of the $W$ and $Z$ bosons at the CERN $\bar{p} p$ collider'',
  Z.\ Phys.\ C {\bf 49}, 17 (1991).

\bibitem{Alitti:1993pn} 
  J.~Alitti {\it et al.}  [UA2 Collaboration],
  ``A search for new intermediate vector mesons and excited quarks decaying to two jets at the CERN $\bar{p} p$ collider'',
  Nucl.\ Phys.\ B {\bf 400}, 3 (1993).


\bibitem{Albajar:1988rs} 
  C.~Albajar {\it et al.}  [UA1 Collaboration],
  ``Two jet mass distributions at the CERN proton - anti-proton collider,''
  Phys.\ Lett.\ B {\bf 209}, 127 (1988).
  

\bibitem{Abe:1989gz} 
  F.~Abe {\it et al.}  [CDF Collaboration],
  ``The two jet invariant mass distribution at $\sqrt{s} = 1.8$ TeV,''
  Phys.\ Rev.\ D {\bf 41}, 1722 (1990).

\bibitem{Abe:1993it}
  F.~Abe {\it et al.}  [CDF Collaboration],
  ``Search for quark compositeness, axigluons and heavy particles using the dijet invariant mass spectrum observed in $p\bar{p}$ collisions,''
  Phys.\ Rev.\ Lett.\  {\bf 71} (1993) 2542.

\bibitem{Abe:1995jz} 
  F.~Abe {\it et al.}  [CDF Collaboration],
  ``Search for new particles decaying to dijets in $p\bar{p}$ collisions at $\sqrt{s} = 1.8$ TeV,''
  Phys.\ Rev.\ Lett.\  {\bf 74}, 3538 (1995)
  [hep-ex/9501001].
  
\bibitem{Abe:1997hm} 
  F.~Abe {\it et al.}  [CDF Collaboration],
  ``Search for new particles decaying to dijets at CDF,''
  Phys.\ Rev.\ D {\bf 55}, 5263 (1997)
  [hep-ex/9702004].

\bibitem{Aaltonen:2008dn} 
  T.~Aaltonen {\it et al.}  [CDF Collaboration],
  ``Search for new particles decaying into dijets in proton-antiproton collisions at $\sqrt{s} = 1.96$ TeV,''
  Phys.\ Rev.\ D {\bf 79}, 112002 (2009)
  [arXiv:0812.4036]. 


\bibitem{Abazov:2003tj} 
  V.~M.~Abazov {\it et al.}  [D0 Collaboration],
  ``Search for new particles in the two jet decay channel", 
  Phys.\ Rev.\ D {\bf 69}, 111101 (2004)
  [hep-ex/0308033].


\bibitem{Aad:2010ae} 
  G.~Aad {\it et al.}  [ATLAS Collaboration],
  ``Search for new particles in two-jet final states in 7 TeV proton-proton collisions", 
  Phys.\ Rev.\ Lett.\  {\bf 105}, 161801 (2010)
  [arXiv:1008.2461]. 
  
\bibitem{Aad:2011aj} 
  G.~Aad {\it et al.}  [ATLAS Collaboration],
  ``Search for New Physics in Dijet Mass and Angular Distributions in pp Collisions at $\sqrt{s} = 7$ TeV", 
  New J.\ Phys.\  {\bf 13}, 053044 (2011)
  [arXiv:1103.3864]. 
  
 \bibitem{ATLAS-CONF-2011-081}
ATLAS Collaboration, "Update of the Search for New Physics in the Dijet Mass
                      Distribution in 163 pb$^{-1}$ of $pp$ Collisions at $\sqrt{s} = 7$
                      TeV", 
note ATLAS-CONF-2011-081, June 2011.

 \bibitem{ATLAS-CONF-2011-095}
ATLAS Collaboration, "Search for New Physics in Dijet Mass Distributions in 0.81
                      fb$^{-1}$ of $pp$ Collisions at $\sqrt{s} = 7$ TeV",  
note ATLAS-CONF-2011-095, July 2011.

\bibitem{Aad:2011fq} 
  G.~Aad {\it et al.}  [ATLAS Collaboration],
  ``Search for new physics in the dijet mass distribution using 1 $fb^{-1}$ of $pp$ collision data at $\sqrt{s}=7$ TeV", 
  Phys.\ Lett.\ B {\bf 708}, 37 (2012)
  [arXiv:1108.6311]. 

\bibitem{ATLAS:2012pu} 
  G.~Aad {\it et al.}  [ATLAS Collaboration],
  ``Search for new phenomena in dijet mass and angular distributions using $pp$ collisions at $\sqrt{s}=7$ TeV,''
  JHEP {\bf 1301}, 029 (2013)
  [arXiv:1210.1718]. 

 \bibitem{ATLAS-CONF-2012-088}
ATLAS Collaboration, "Search for New Phenomena in the Dijet Mass Distribution
                      using 5.8 $fb^{-1}$ of $pp$  Collisions at $\sqrt{s}=8$ TeV", 
note ATLAS-CONF-2012-088, July 2012.

\bibitem{ATLAS-CONF-2012-148}
ATLAS Collaboration, "Search for New Phenomena in the Dijet Mass Distribution
updated using 13.0 fb$^{-1}$ of $pp$ Collisions at $\sqrt{s}=8$ TeV",
note ATLAS-CONF-2012-148, Nov. 2012.


\bibitem{Khachatryan:2010jd} 
  V.~Khachatryan {\it et al.}  [CMS Collaboration],
  ``Search for Dijet Resonances in 7 TeV pp Collisions at CMS,''
  Phys.\ Rev.\ Lett.\  {\bf 105}, 211801 (2010)
  [arXiv:1010.0203 [hep-ex]].

\bibitem{Chatrchyan:2011ns} 
  S.~Chatrchyan {\it et al.}  [CMS Collaboration],
  ``Search for Resonances in the Dijet Mass Spectrum from 7 TeV pp Collisions at CMS,''
  Phys.\ Lett.\ B {\bf 704}, 123 (2011)
  [arXiv:1107.4771]. 

\bibitem{CMS:2012yf} 
  S.~Chatrchyan {\it et al.}  [CMS Collaboration],
  ``Search for narrow resonances and quantum black holes in inclusive and $b$-tagged dijet mass spectra from $pp$ collisions at $\sqrt{s}=7$ TeV,''
  JHEP {\bf 1301}, 013 (2013)
  [arXiv:1210.2387]. 

\bibitem{CMS-PAS-EXO-11-094}
CMS Collaboration, "Search for narrow resonances using the dijet mass spectrum in pp collisions at $\sqrt{s} = 7$ TeV", note CMS-PAS-EXO-11-094, July 2012.

\bibitem{Chatrchyan:2013qha} 
  S.~Chatrchyan {\it et al.}  [CMS Collaboration],
  ``Search for narrow resonances using the dijet mass spectrum in pp collisions at $\sqrt{s} = 8$ TeV,''
  arXiv:1302.4794 [hep-ex].

\bibitem{CMS-PAS-EXO-12-059}
CMS Collaboration, "Search for Narrow Resonances using the Dijet Mass Spectrum
with 19.6 fb$^{-1}$ of pp Collisions at $\sqrt{s}=8$ TeV", note CMS-PAS-EXO-12-059, Feb. 2013.


\bibitem{Bauer:2009cc} 
  C.W.~Bauer, {\it et al}, 
  ``Supermodels for early LHC,''
  Phys.\ Lett.\ B {\bf 690}, 280 (2010)
  [arXiv:0909.5213]. 

\bibitem{Manohar:2006ga} 
  A.~V.~Manohar and M.~B.~Wise,
  ``Flavor changing neutral currents, an extended scalar sector, and the Higgs production rate at the LHC,''
  Phys.\ Rev.\ D {\bf 74}, 035009 (2006)
  [hep-ph/0606172].

\bibitem{Babu:1996vt} 
  K.~S.~Babu, C.~F.~Kolda and J.~March-Russell,
  ``Leptophobic U(1) $s$ and the R($b$) - R($c$) crisis,''
  Phys.\ Rev.\ D {\bf 54}, 4635 (1996)
  [hep-ph/9603212].

\bibitem{Georgi:1996ei} 
  H.~Georgi and S.~L.~Glashow,
  ``Decays of a leptophobic gauge boson,''
  Phys.\ Lett.\ B {\bf 387}, 341 (1996)
  [hep-ph/9607202].


\bibitem{Rosner:1996eb} 
  J.~L.~Rosner,
  ``Prominent decay modes of a leptophobic $Z^\prime$,''
  Phys.\ Lett.\ B {\bf 387}, 113 (1996)
  [hep-ph/9607207].

\bibitem{FileviezPerez:2011pt} 
  P.~Fileviez Perez and M.~B.~Wise,
  ``Breaking local baryon and lepton number at the TeV scale,''
  JHEP {\bf 1108}, 068 (2011)
  [arXiv:1106.0343]. 

\bibitem{Hill:1991at}
  C.~T.~Hill,
  ``Topcolor: Top quark condensation in a gauge extension of the standard model,''
  Phys.\ Lett.\  {\bf B266}, 419-424 (1991) ; \\
  C.~T.~Hill, S.~J.~Parke,
  ``Top production: Sensitivity to new physics,''
  Phys.\ Rev.\  {\bf D49}, 4454-4462 (1994).
  [hep-ph/9312324].

\bibitem{Chivukula:1996yr}
  R.~S.~Chivukula, A.~G.~Cohen, E.~H.~Simmons,
  ``New strong interactions at the Tevatron?,''
  Phys.\ Lett.\  {\bf B380}, 92-98 (1996).
  [hep-ph/9603311];

  E.~H.~Simmons,
  ``Coloron phenomenology,''
  Phys.\ Rev.\  {\bf D55}, 1678-1683 (1997).
  [hep-ph/9608269].

\bibitem{Bai:2010dj} 
  Y.~Bai and B.~A.~Dobrescu,
  ``Heavy octets and Tevatron signals with three or four b jets,''
  JHEP {\bf 1107}, 100 (2011)
  [arXiv:1012.5814].

\bibitem{Chivukula:2013kw} 
  R.~S.~Chivukula, E.~H.~Simmons and N.~Vignaroli,
  ``A Flavorful Top-Coloron Model,''
  Phys.\ Rev.\ D {\bf 87}, 075002 (2013)
  [arXiv:1302.1069];
  ``Same-sign dileptons from colored scalars in the Flavorful Top-Coloron Model'', arXiv:1306.2248.

\bibitem{Kumar:2012ww} 
  K.~Kumar, R.~Vega-Morales and F.~Yu,
  ``Effects from New Colored States and the Higgs Portal on Gluon Fusion and Higgs Decays,''
  Phys.\ Rev.\ D {\bf 86}, 113002 (2012)
  [arXiv:1205.4244].

\bibitem{Batra:2005rh} 
  P.~Batra, B.~A.~Dobrescu and D.~Spivak,
  ``Anomaly-free sets of fermions,''
  J.\ Math.\ Phys.\  {\bf 47}, 082301 (2006)
  [hep-ph/0510181].


\bibitem{Carena:2004xs} 
  M.~S.~Carena, A.~Daleo, B.~A.~Dobrescu and T.~M.~P.~Tait,
  ``$Z^\prime$ gauge bosons at the Tevatron,''
  Phys.\ Rev.\ D {\bf 70}, 093009 (2004)
  [hep-ph/0408098]. \ 
  M.~Duerr, P.~Fileviez Perez and M.~B.~Wise,
  ``Gauge Theory for Baryon and Lepton Numbers with Leptoquarks,''
  Phys.\ Rev.\ Lett.\  {\bf 110}, 231801 (2013)
  [arXiv:1304.0576 [hep-ph]].

\bibitem{Carone:1994aa} 
  C.~D.~Carone and H.~Murayama,
  ``Possible light U(1) gauge boson coupled to baryon number,''
  Phys.\ Rev.\ Lett.\  {\bf 74}, 3122 (1995)
  [hep-ph/9411256];
  ``Realistic models with a light U(1) gauge boson coupled to baryon number,''
  Phys.\ Rev.\ D {\bf 52}, 484 (1995)
  [hep-ph/9501220]. \\
  P.~Fileviez Perez and M.~B.~Wise,
  ``Baryon and lepton number as local gauge symmetries,''
  Phys.\ Rev.\ D {\bf 82}, 011901 (2010)
  [Erratum-ibid.\ D {\bf 82}, 079901 (2010)]
  [arXiv:1002.1754].
 

\bibitem{Chatrchyan:2013sfs} 
  S.~Chatrchyan {\it et al.}  [CMS Collaboration],
  ``Searches for Higgs bosons in pp collisions at sqrt(s) = 7 and 8 TeV in the context of four-generation and fermiophobic models,''
  Phys.\ Lett.\ B {\bf 725}, 36 (2013)
  [arXiv:1302.1764 [hep-ex]].
  ATLAS Collaboration,
 ``Update of the Combination of Higgs Boson Searches in pp Collisions at $\sqrt{s} = 7$ TeV,''
note ATLAS-CONF-2011-135.



\bibitem{tprime}
  G.~Aad {\it et al.}  [ATLAS Collaboration],
  ``Search for pair production of heavy top-like quarks decaying to a high-$p_T$ $W$ boson and a $b$ quark in the lepton plus jets final state at $\sqrt{s}=7$ TeV", 
  Phys.\ Lett.\ B {\bf 718}, 1284 (2013)
  [arXiv:1210.5468]; 
\\
S.~Chatrchyan {\it et al.}  [CMS Collaboration],
  ``Search for heavy, top-like quark pair production in the dilepton final state in $pp$ collisions at $\sqrt{s} = 7$ TeV,''
  Phys.\ Lett.\ B {\bf 716}, 103 (2012)
  [arXiv:1203.5410];   
\
  ``Search for pair produced fourth-generation up-type quarks in $pp$ collisions at $\sqrt{s}=7$ TeV with a lepton in the final state,''
  Phys.\ Lett.\ B {\bf 718}, 307 (2012)
  [arXiv:1209.0471].

\bibitem{bprime}
  S.~Chatrchyan {\it et al.}  [CMS Collaboration],
 ``Search for heavy bottom-like quarks in 4.9 inverse femtobarns of $pp$ collisions at $\sqrt{s}=7$ TeV,''
  JHEP {\bf 1205}, 123 (2012)
  [arXiv:1204.1088]; 
\
  ``Search for heavy quarks decaying into a top quark and a $W$ or $Z$ boson using lepton + jets events in $pp$ collisions at $\sqrt{s}=7$ TeV,''
  JHEP {\bf 01}, 154 (2013)
  [arXiv:1210.7471];
  ``Combined search for the quarks of a sequential fourth generation,''
  Phys.\ Rev.\ D {\bf 86}, 112003 (2012)
  [arXiv:1209.1062].

   
\bibitem{Hall:1985wz} 
  L.~J.~Hall and A.~E.~Nelson,
  ``Heavy Gluons And Monojets,''
  Phys.\ Lett.\ B {\bf 153}, 430 (1985); \
P.~H.~Frampton and S.~L.~Glashow,
  ``Chiral Color: An Alternative to the Standard Model,''
Phys.\ Lett.\ B {\bf 190}, 157 (1987).
  
\bibitem{Dobrescu:2009vz} 
  B.~A.~Dobrescu, K.~Kong and R.~Mahbubani,
  ``Prospects for top-prime quark discovery at the Tevatron,''
  JHEP {\bf 0906}, 001 (2009)
  [arXiv:0902.0792].
 
  
\bibitem{Dobrescu:2007yp}
  B.~A.~Dobrescu, K.~Kong and R.~Mahbubani,
  ``Massive color-octet bosons and pairs of resonances at hadron colliders,''
  Phys.\ Lett.\  B {\bf 670}, 119 (2008)
  [arXiv:0709.2378].

\bibitem{Quigg:2009gg} 
  C.~Quigg,
  ``LHC Physics Potential versus Energy,''
  arXiv:0908.3660 [hep-ph].

\bibitem{CMS:2012ooa} 
  S.~Chatrchyan {\it et al.}  [CMS Collaboration],
  ``Data Parking and Data Scouting at the CMS Experiment,''
  CMS-DP-2012-022.

\bibitem{Chivukula:2011ng} 
  R.~S.~Chivukula, A.~Farzinnia, E.~H.~Simmons and R.~Foadi,
  ``Production of massive color-octet vector bosons at Next-to-Leading Order,''
  Phys.\ Rev.\ D {\bf 85}, 054005 (2012)
  [arXiv:1111.7261]. \
  R.~S.~Chivukula, A.~Farzinnia, J.~Ren and E.~H.~Simmons,
  ``Hadron Collider Production of Massive Color-Octet Vector Bosons at Next-to-Leading Order,''
  Phys.\ Rev.\ D {\bf 87}, 094011 (2013)
  [arXiv:1303.1120 [hep-ph]].
  

\bibitem{Alwall:2011uj} 
  J.~Alwall, {\it et al}, 
  ``MadGraph 5 : Going Beyond,''
  JHEP {\bf 1106}, 128 (2011)
  [arXiv:1106.0522]. 

\bibitem{Pumplin:2002vw} 
 J.~Pumplin, {\it et al}, 
  ``New generation of parton distributions with uncertainties from global QCD analysis,''
  JHEP {\bf 0207}, 012 (2002)
  [hep-ph/0201195].

\bibitem{Sjostrand:2006za} 
 T.~Sjostrand, S.~Mrenna and P.~Z.~Skands,
  ``PYTHIA 6.4 Physics and Manual,''
  JHEP {\bf 0605}, 026 (2006)
  [hep-ph/0603175].

\bibitem{PGS4}
  J.~Conway {\it et al}, 
  ``Pretty Good Simulation of high energy collisions",
   {\tt http://physics.ucdavis.}\\  {\tt edu/$\small \sim$conway/research/software/pgs}\\{\tt /pgs4-general.htm}

\bibitem{Krnjaic:2011ub} 
  G.~Z.~Krnjaic,
  ``Very light axigluons and the top asymmetry,''
  Phys.\ Rev.\ D {\bf 85}, 014030 (2012)
  [arXiv:1109.0648]. \\
  M.~Gresham, J.~Shelton and K.~M.~Zurek,
  ``Open windows for a light axigluon explanation of the top forward-backward asymmetry,''
  JHEP {\bf 1303}, 008 (2013)
  [arXiv:1212.1718].


\bibitem{Yu:2011cw} 
  F.~Yu,
  ``A $Z'$ model for the CDF dijet anomaly,''
  Phys.\ Rev.\ D {\bf 83}, 094028 (2011)
  [arXiv:1104.0243]. 
  
\bibitem{Choudhury:2011cg} 
  D.~Choudhury, R.~M.~Godbole and P.~Saha,
  ``Dijet resonances, widths and all that,''
  JHEP {\bf 1201}, 155 (2012)
  [arXiv:1111.1054 [hep-ph]].
  
\bibitem{Atre:2012gj} 
  A.~Atre, {\it et al}, 
  ``Probing color octet couplings at the LHC'',
  Phys.\ Rev.\ D {\bf 86}, 054003 (2012)
  [arXiv:1206.1661].
  
\bibitem{Aaltonen:2009qu} 
  T.~Aaltonen {\it et al.}  [CDF Collaboration],
  ``Search for the production of narrow $t\bar{b}$ resonances in 1.9 fb$^{-1}$ of $p\bar{p}$ collisions at $\sqrt{s} = 1.96$ TeV,''
  Phys.\ Rev.\ Lett.\  {\bf 103}, 041801 (2009)
  [arXiv:0902.3276].

\bibitem{Zerwekh:2009vi} 
  A.~R.~Zerwekh,
  ``Axigluon Couplings in the Presence of Extra Color-Octet Spin-One Fields,''
  Eur.\ Phys.\ J.\ C {\bf 65}, 543 (2010)
  [arXiv:0908.3116 [hep-ph]]. \
  ``Constraining Spin-One Color-Octet Resonances Using CDF and ATLAS Data,''
  Eur.\ Phys.\ J.\ C {\bf 70}, 917 (2010)
  [arXiv:1008.4575 [hep-ph]].
  
\bibitem{Gaiser:1982yw} 
  J.~Gaiser,
  ``Charmonium spectroscopy from radiative decays of the $J / \psi$ and $\psi^\prime$,''
  SLAC-255 (1983), see Appendix F.1.

\end{thebibliography}
\end{document}